\documentclass[AER]{AEA}

\usepackage{natbib}

\draftSpacing{1.5}

\usepackage{fdsymbol}
\usepackage{graphicx}
\usepackage{url}
\usepackage[utf8]{inputenc}

\graphicspath{ {./images/} }

\begin{document}

\title{Economic Performance Through Time: \\ A Dynamical Theory}
\shortTitle{Economic Performance Through Time}
\author{Daniel Seligson and Anne McCants\thanks{Seligson: Unaffiliated, 1741 Middlefield Rd., Palo Alto, CA 94301, daniel.seligson@gmail.com. 
McCants: MIT, Dept. of History, E51-263, 77 Mass. Ave., Cambridge, MA 02139.}}
\date{\today}
\pubMonth{May}
\pubYear{2019}
\pubVolume{}
\pubIssue{}

\def\myD{{\cal D}}
\def\myE{{\cal E}}
\def\myP{{\cal P}}
\def\myI{{\cal I}}
\def\myN{{\cal N}}
\def\myO{{\cal O}}
\def\myFP{\phi^i_0}

\begin{abstract}
The central problems of Development Economics are the explanation of
the gross disparities in the global distribution, $\myD$, of economic
performance, $\myE$, and its persistence, $\myP$. Douglass North argued,
epigrammatically, that institutions, $\myI$, are the rules of the game,
meaning that $\myI$ determines or at least constrains $\myE$. This promised to
explain $\myD$. $65,000$ citations later, the central problems remain
unsolved. North’s institutions, $\myI_N$, are informal, slowly changing
cultural norms as well as roads, guilds, and formal legislation that
may change overnight. This definition, mixing the static and the
dynamic, is unsuited for use in a necessarily time-dependent theory of
developing economies. We offer here a suitably precise definition of
$\myI$, a dynamical theory of economic development, a new measure of the
economy, an explanation of $\myP$, a bivariate model that explains half of
$\myD$, and a critical reconsideration of North’s epigram.
\end{abstract}

\maketitle

%

Why do the haves have and the have-nots haven't? This is one of the
two central questions of Development Economics and Economics
History. The Nobelist Douglass North opined in 1990 that, despite $40$
years of immense effort, there had been scant progress answering
it \citep{North1990}. The lesser appreciated question concerns the persistence of
having. Though his monograph on the role of institutions on economic
performance through time was triumphal in tone and is now the
14$^{\rm th}$ most cited book in all the social sciences, we may say in
2019 that after $70$ years of immense effort, progress on either
question remains scant.

We present a theory of economic development that begins with a new translation of North’s narrative into mathematics. A theory of
development must address growth and loss, not only in the objective
function, economic performance, but in its source terms, in this case,
institutions. A differential equations approach is well-suited to
this. We find that a 2-dimensional linear dynamical theory of
Development Economics has verisimilitude, 
refines our understanding of institutions and their role, and
overturns convention. The task of answering why the haves have demands
more space than allotted here, but we show a two-factor model that
accounts for half of the global variance, and thus we half explain why
the haves have.

\section{The Persistence of Fortune}

The disparities in economic performance, $\myE$, are stark. Aggregating at the national level, per capita Gross National Income in $2017$ \citep{GNI2017} ranged from about $\$128k$ in Qatar to $\$730$ in Burundi, that is, by $175\!:\!1$. Economic historians use the term “The Great Divergence” to encapsulate the temporal growth of that range, and much research has been devoted to its origins. \citep{Pomeranz2001} We ask instead, what may be said about the relative economic fortunes of nations over the course of time, and what, if anything, does that tell us about why the world’s haves have and why their having is persistent, these being the two central and unsolved problems of Development Economics and Economic History according to North. \citep{North1990}

Let the UN’s Human Development Index (HDI) serve as a proxy for $\myE$. Escosura and co-workers \citep{Escosura2015} have estimated HDI for $164$ countries at $24$ dates covering the period $1870$ to $2015$, and a UN archive includes annual data from $1990$ to the present for $190$ countries \citep{UNHDI}. We denote the economic performance for a nation $i$ in year $t$ as $\myE^i(t)$ and the distribution over all $i$ as $\myD(t)$ . Let the autocorrelation function of $\myD$  between $t$ and $t-\tau$ be the persistence, $\myP_{t,\tau}$. It is insensitive to temporal changes in scale, e.g., $\myD(t)=f(\tau)*\myD(t-\tau)$, while it remains sensitive to changes in relative position of the $\myE^i$ arising from, e.g., stochastic rank order changes or a systematic nonlinear expansion of the $\myD$ universe.

\begin{figure} 
\includegraphics[width=9cm, height=6cm]{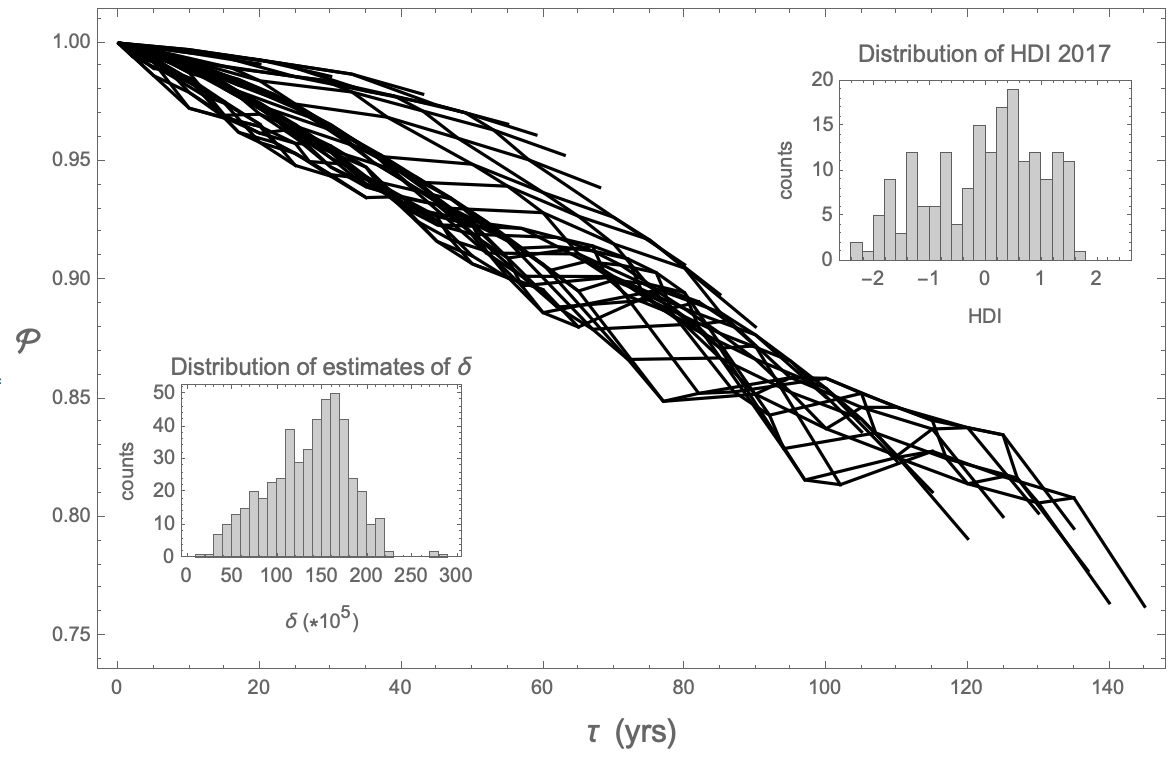}
\caption{Persistence in the global economy.}
\label{fig:Escosura}
\begin{figurenotes}
$\myP$, the autocorrelation function of the distribution of indices of human development from 1870 to 2015 \citep{Escosura2015} shows slow, steady erosion. Lower inset: The distribution of year-to-year erosions extracted from \citep{Escosura2015} and \citep{UNHDI}, implying that the half-life of  is approximately 500 years. Upper inset: Contemporary HDI is zero-centered unimodal.
\end{figurenotes}
\end{figure}

As illustrated in Fig. \ref{fig:Escosura} for all $t$ and $\tau$ in the long term dataset, the persistence erodes continuously, slowly, and independently of $t$, falling only $23\%$ after $145$ years. 
This suggests an approximation of the form
\begin{equation} \nonumber
\myP_{t,\tau}=(1-\delta)^\tau
\end{equation}
where $\delta$ is the year-to-year loss in persistence. Employing all $N=486$ unique correlations among the two datasets, we estimate $\delta=0.00135 \pm 0.00045$. We may now answer our earlier question, “What may be said about the relative economic fortunes of nations?” Specified in units such that $\bar{\myD}=0$ and $\sigma^2_\myD=1$, $\myD$ is materially constant over the course of a century. Its half-life is half a millennium.

\section{A Dynamical Theory of Institutions and Economic Performance} \label{exposition}

In (1), North elaborated on his epigram, Institutions are the rules of the game.
\begin{quotation}
“Institutions are the humanly devised constraints that structure political, economic, and social interaction. They consist of both informal constraints (sanctions, taboos, customs, traditions, and codes of conduct), and formal rules (constitutions, laws, and property rights). Throughout history, institutions have been devised by human beings to create order and reduce uncertainty in exchange. Together with the standard constraints of economics they define the choice set and therefore determine the transaction and production costs and hence the profitability and feasibility of engaging in economic activity. They evolve incrementally, connecting the past with the future; history in consequence is largely a story of institutional evolution in which the historical performance of economies can only be understood as a part of a sequential story.”
\end{quotation}
Rules of the game is a play on words, marrying a colloquial meaning, how it is done, with North’s avowed preference for game theory. But his persuasive prose suggests differential equations, not game theory.

Robinson et al. \citep{AJR2001} were among the first to codify North in mathematical terms. They wrote,
\begin{equation} \label{eq:AJR}
\myE^i=\alpha\times \myI^i_N+f(\vec{x}^i)
\end{equation}
where $\myI_N$ is some measure of institutions, $\alpha$ is a positive constant, and $f$ is a differentiable function over factors $\vec{x}$ that are broad enough in concept to include what North called the standard constraints of economics, for instance climate, geography, and natural resources.  In all cases, the superscript $i$ is the polity-level index. Eq. \ref{eq:AJR} falls short if only because it lacks the one ingredient essential to any theory of history or development, that is, time.

Later, Dell et al. \citep{DJO2009} wrote 
\begin{equation} \label{eq:DJO}
\dot{\myE}^i=\alpha\times \myI^i_N+f(\vec{x}^i)
\end{equation}
If Eq. \ref{eq:AJR} states that institutions beget wealth, Eq. \ref{eq:DJO} states that institutions beget economic growth. Inclusion of the time derivative is an improvement, but North points out that $\myI_N$ and $\myE$ evolve incrementally. Thus, we need a complementary process, such as
\begin{equation} \label{eq:Complement}
\dot{\myI}^i_N=\beta\times \myE^i +g(\vec{x}^i) 
\end{equation}
which encodes the widely acknowledged if overlooked fact that institutions are expensive. The linear scale transformation $\myI_N \rightarrow \sqrt{\beta/\alpha}*\myI_N$ makes the coefficients of $\myI_N$ and $\myE$ identical, so without loss of generality, we may write $\beta=\alpha$. That coefficient, however we write it, couples $\myI_N$ and $\myE$ but does not constrain them. This is not merely a semantic point.

\begin{figure}
\includegraphics[width=6cm, height=6cm]{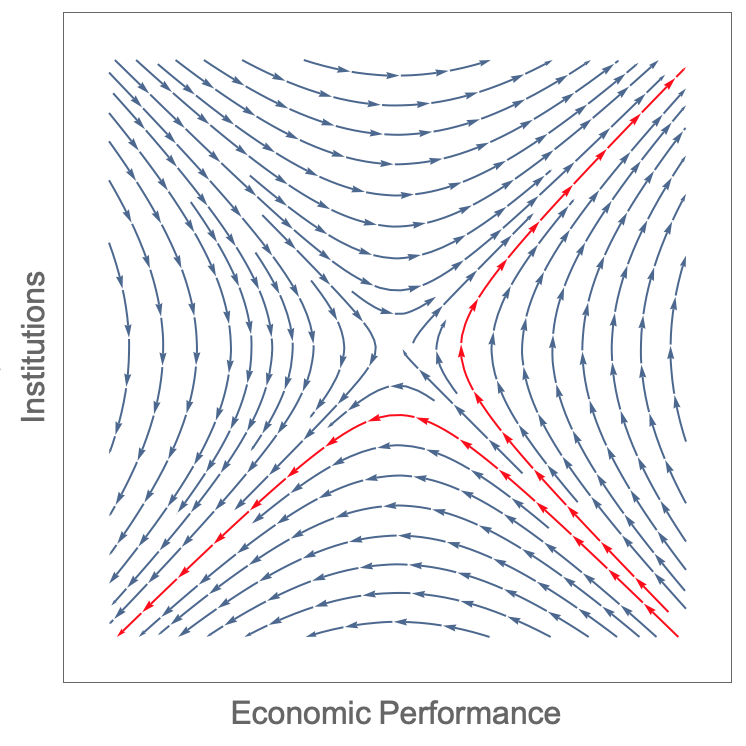}
\caption{Unstable Streamflows}
\label{fig:UnstableFlows}
\begin{figurenotes}
Streamlines of the unstable trajectory of a polity in a phase space governed by Eqs. \ref{eq:DJO} and \ref{eq:Complement}. Without loss of generality, we place the unstable fixed point at the origin. Wherever a polity begins its journey, it is swept toward the positive sloping diagonal, the line $\myI=\myE$, and then accelerated away to $\pm\infty$ along that diagonal. Outcomes are extremely sensitive to initial conditions, as illustrated by the red streamlines, dependent only on whether its initial position was above or below the negative sloping diagonal, the line $\myI=-\myE$.
\end{figurenotes}
\end{figure}

Eqs. \ref{eq:DJO} and \ref{eq:Complement} describe an unconstrained dynamical system. In the phase space defined by $\myI_N$ and $\myE$, the system has a saddle point, an unstable fixed point, shown in Fig. \ref{fig:UnstableFlows} 

The streamlines of the flow drive a state toward the northeast-pointing diagonal, thereupon ejecting it to $\pm \infty$ along that axis. The ejection proceeds exponentially at a rate determined by the coupling, $\alpha$. Consequently, a developed global economy would have exponentially rich states and exponentially poor ones and no center. This would be reflected in a hollowing out of $\myD$. But $\myD$ is zero-centered and unimodal, as is seen in the distribution of $2017$ HDI in Fig. 1, and persistently so. Thus Eqs. \ref{eq:DJO} and \ref{eq:Complement} are at best incomplete.

Furthermore, consider homophilous or like-states $i$ and $j$ such that $\vec{x}^i \!\simeq \!\vec{x}^j$ for instance Arabian Gulf States, Scandinavia, or the Balkans. Like-states have evolved together and $\myE^i \!\simeq \!\myE^j$. Yet the solutions of Eqs. \ref{eq:DJO} and \ref{eq:Complement} may eject them to opposite corners of the economic universe in Fig. 2, their fates determined only by arbitrarily small differences in their initial conditions. Such homophilies are common, and a dynamical system that does not preserve them lacks verisimilitude. Eqs. \ref{eq:DJO} and \ref{eq:Complement} disappoint on this condition, too. The system is too simple.

The premise of \citep{AJR2001}, \citep{DJO2009}, and others is that what North calls the standard constraints of economics are realized in the exogenous factors. The solutions of Eqs. \ref{eq:DJO} and \ref{eq:Complement} demonstrate that this premise is false. An additional dissipative or drag term is needed to constrain them, or equivalently, to stabilize the fixed point. Proceeding accordingly, we get, in the most general case
\begin{equation} \label{eq:Stable0}
\begin{split}
\dot{\myE}^i & = -\lambda\times \myE^i+\alpha\times \myI^i_N+f(\vec{x}^i) \\
\dot{\myI}^i_N & =\ \ \alpha\times \myE^i  -\gamma\times \myI^i_N + g(\vec{x}^i) 
\end{split}
\end{equation}
Are the drag terms artifice or do they have economic meaning? Consider a state that provides good health care and longevity to its inhabitants. In doing so, it creates a cohort of too-old-to-work citizens who contribute little revenue to the economy. The support of those by others is a drag on growth. This is hardly the only such example. 

Before we discuss the stability of this system, the meaning of its solutions, tests of their validity, and its implications, let us first contemplate $\dot{\myI}^i_N$. Per North, $\myI_N$ are humanly devised informal constraints, e.g., taboos and codes of conduct, and formal rules, e.g., constitutions and laws.  $\myI_N$ is thus a composite, some of whose elements change very slowly, e.g. taboos, and others much more quickly, e.g. laws. We expose its composite nature by writing $\myI_N=\myI+\myN$, where $\myI$ are the more rapidly varying elements, which we shall call Institutions, and $\myN$ are the more slowly varying elements, which we shall call Norms. It follows that $\dot{\myI}\gg\dot{\myN}$ and $\dot{\myI_N}=\dot{\myI}+\dot{\myN}\simeq\dot{\myI}$ . This result ripples through Eq. \ref{eq:Stable0} leading to
\begin{equation} \label{eq:Stable1}
\begin{split}
\dot{\myE}^i & = {-}\lambda\times \myE^i+\alpha\times \myI^i+f(\vec{x}^i,\myN^i) \\
\dot{\myI}^i & =\ \ \alpha\times \myE^i  -\gamma\times \myI^i + g(\vec{x}^i,\myN^i) 
\end{split}
\end{equation}
It must be apparent at this juncture that slowly varying Norms and rapidly varying Institutions, though one and the same colloquially, have very different influence on the ebbs, flows, and equilibria of our dynamical system and must be conceived of and deployed as distinct types.

Eq. \ref{eq:Stable1} is a system of first order linear differential equations whose exponential solutions are well understood, but we will simplify them further. First, we redefine the constants such that $\alpha\rightarrow\alpha/\lambda$ and $\gamma\rightarrow\gamma/\lambda$. Second, we conjecture that the dissipation is more or less equal for both $\myI$ and $\myE$. We will test this later, but for now we set $\gamma=1$, leaving
\begin{equation} \label{eq:Stable2}
\begin{split}
\dot{\myE}^i & = \lambda(-\myE^i+\alpha\times \myI^i) +f(\vec{x}^i,\myN^i) \\
\dot{\myI}^i & =\lambda(\alpha\times \myE^i  \ -\ \myI^i) + g(\vec{x}^i,\myN^i) 
\end{split}
\end{equation}

\begin{figure}
\includegraphics[width=12cm, height=4cm]{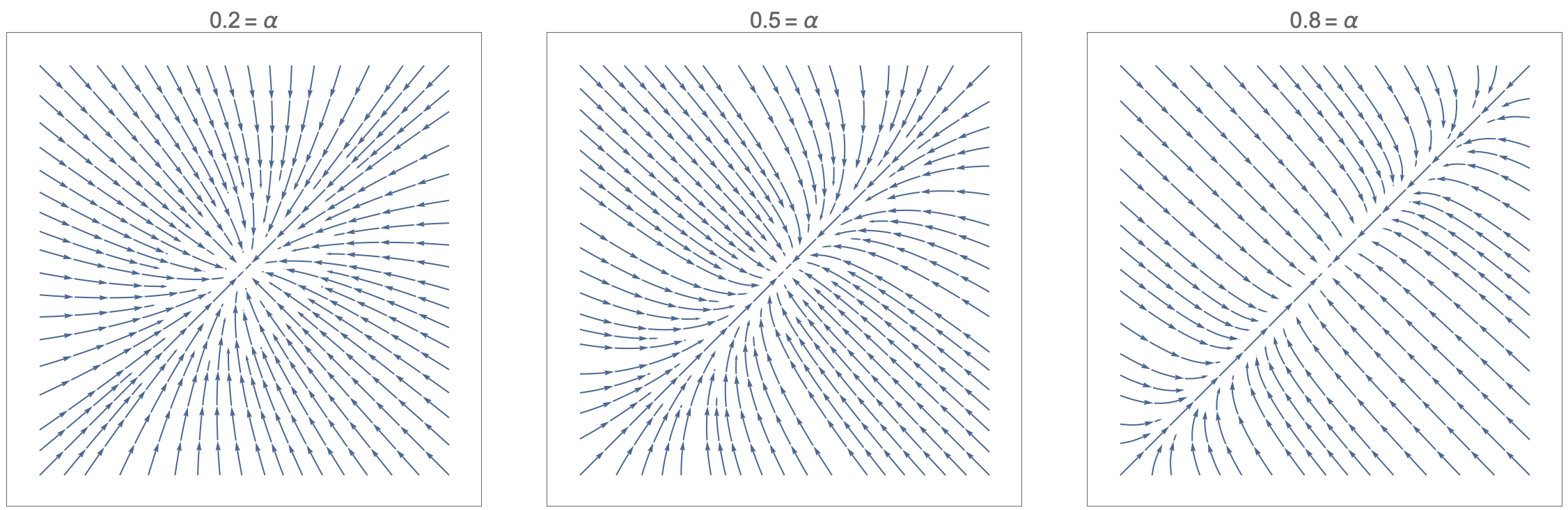}
\caption{Streamlines of the stable trajectory governed by Eq. \ref{eq:Stable2}}
\label{fig:StableFlows}
\begin{figurenotes}
The $\alpha$-dependent flow for $\alpha=0.2,\ 0.5,\  \text{and}\ 0.8$.  $\myI$ is the ordinate and $\myE$ the abscissa in each panel. Irrespective of initial conditions, a polity flows to the system’s fixed point, $\phi_0$, shown here at the origin without loss of generality.
\end{figurenotes}
\end{figure}

The eigenvectors of Eq. \ref{eq:Stable2} are the northeast- and southeast-pointing diagonals $\mu=\myI+\myE$ and $\kappa=\myI-\myE$. The negative inverse of its eigenvalues are the time constants of the exponentially converging solutions,
\begin{equation} \label{eq:Taus}
\tau_{\mu,\kappa}=\frac{1}{\lambda}\frac{1}{1\mp\alpha}
\end{equation}
For $0\le\alpha<1$, both time constants are positive, and irrespective of its initial condition, a state $i$ flows to a fixed point $\myFP$. The flow is $\alpha$-dependent, as may be seen in Fig. \ref{fig:StableFlows}. 
The relative dissipation strength, $\gamma$, which we have set to unity, determines the slope of the primary diagonal.

\section{Are Institutions the Rules of the Game?}

Consider the locus of fixed points, $\myFP$, in the coordinate system of the eigenvectors, where $f^i$ denotes $f(\vec{x}^i,\myN^i)$ and so forth.
\begin{equation} \label{eq:FixedPoints}
\phi^i_0=\left\{\mu^i_0,\kappa^i_0 \right\}=\frac{1}{\lambda}\left\{\frac{f^i+g^i}{1-\alpha},\frac{f^i-g^i}{1+\alpha}\right\}
\end{equation}

All that we have said thus far about $\vec{x}$ is that it is likely to be affected by climate, geography, and natural resources.  No doubt, the acute events of history, e.g., war and natural disaster, contribute, too. Climate and geography, like Norms, change only over centuries. If these dominate $f$ and $g$, then each fixed point is more or less fixed over centuries, and this would explain the semi-millennial persistence of $\myD$. Also, differentiability of $f$ and $g$ guarantees that the fixed points of like-states are themselves close. Thus the dynamical system preserves homophilies, too.

Eq. \ref{eq:FixedPoints} also informs us that $\myFP$ is determined by the coupling and the drag, and by the arguments $\vec{x}^i$ and $\myN^i$  through the mediations of $f$ and $g$, but not by the Institutions. This arises from the structure of Eq. \ref{eq:Stable2} whose bi-directional coupling of $\myI$ to $\myE$ dictates that equilibrium lies in a corresponding phase space. Both $\myI$ and $\myE$ are dependent variables, a finding that contradicts North’s epigram. As we have split North’s $\myI_N$ into two pieces, is that contradiction a triviality of naming conventions? 

In $2016$, developed countries provided $\$180B$ in foreign assistance to lesser developed countries. Of this, at least  $\$37B$ was committed to infrastructure, governance, security, and building civil society \citep{UnbundlingAid}, that is, to $\myI$ and not to $\myN$, with the expectation that these infusions from without, $\myI_x$, would lead promptly to greater $\myE$. However, in as much as $\myFP$ is dictated by $\myN$, which is more or less constant because it is less malleable than $\myI$ in general, and particularly so from without, it follows from the dynamics that these $\myI_x$ induce an equal and opposite change in endogenous institutions, leaving the sum unchanged, and $\myE$  unchanged, too.

In other words, to effect a change $\Delta\myI$, one must effect a change $\Delta\myN$ as dictated by Eq. \ref{eq:FixedPoints}. Norms are, by definition, resistant to change. While it is not theoretically impossible to effect a $\Delta\myN$, it is practically impossible on the short time scale hoped for by those providing assistance.  Finally, though assertions that $\Delta\myI$ induce $\Delta\myN$ are not refutable, indirect inducement can be neither more prompt nor more efficient than direct intervention.

Summarizing these ideas, North’s epigram deploys a definition of institutions too broad for dynamical systems theory or for quantitative analysis. We parse the definition, deploy it in the mechanism he lays out in the block quote opening Section \ref{exposition}, and find that the epigram does not hold. Without qualification, we may say that Institutions, $\myI$, are not the rules of the game. Norms, $\myN$, and the exogenous $\vec{x}$ determine the endgame.

\section{More Quantitative Tests}

If the coupling of $\myI$ to $\myE$  is very strong,  $\alpha \!\simeq \!1$ and $\tau_\mu \rightarrow \infty$ per Eq. \ref{eq:Taus}. We invest in the judiciary to protect property rights and we invest in roads to facilitate commerce because we expect both to have an impact within years, perhaps decades, possibly even generations, but not $\infty$.  Therefore, $\alpha \!\simneqq\! 1$. On the other hand, $\alpha \! \simneqq \! 0$ because weakly coupled Institutions do not promote economic growth and are of no concern here. We conjecture, then, that 
\begin{equation} \nonumber
\alpha=\myO(0.5)
\end{equation} 
For purposes of estimation we use $\alpha=0.5$.

The constraint $\lambda$ is the drag on growth whose influence grows in concert with $\myI$ or $\myE$. It is the cost of doing business and of systemic waste. Its units are inverse time. North estimates that more than $40\%$ of spending in modern economies is devoted to building and maintaining institutions, that is, to the cost of doing business. We take $\lambda=0.2$ as a working estimate. Eq. \ref{eq:Taus} then informs us that the time constants for convergence to equilibrium are $\tau_\mu=10$ and $\tau_\kappa=3.3$, both measured in years, and both consistent with experience.

Let us now consider the functions $f$ and $g$, both of which take the same arguments. We assert that more or less the same considerations drive $\dot{\myE}$  and $\dot{\myI}$, and therefore $f\simeq g$. From Eq. \ref{eq:FixedPoints}, it follows that on average $\kappa^i_0 \ll \mu^i_0$, or
\begin{equation}\label{eq:Variances}
\frac {\sigma^2_{\mu_0}}{\sigma^2_{\kappa_0}} = \left ( \frac {1+\alpha}{1-\alpha} \right )^2
\end{equation}
where the variances are taken over all states $i$. For $\alpha=0.5$, the ratio is $9$, implying that the equilibrium distribution should have a large principal component along the $\mu=\myI+\myE$ axis, making $\mu$ the proper measure of Development Economics. 

\begin{figure}
\includegraphics[width=10cm, height=10cm]{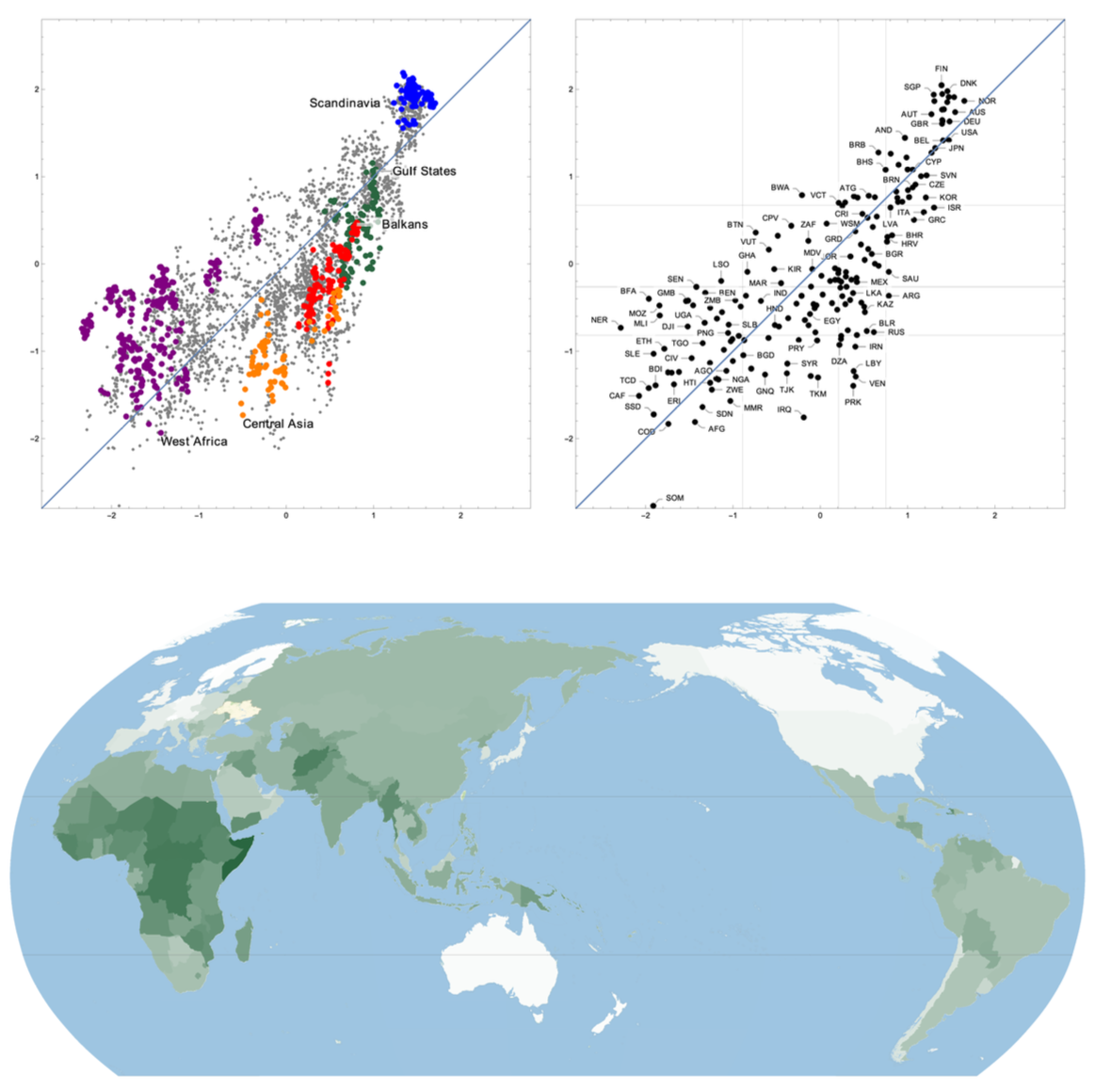}
\caption{Contemporary Institutions and Economic Performance.}
\label{fig:RawData}
\begin{figurenotes}
Upper Left: 3210 joint observations of WGI (ordinate) and HDI (abscissa) covering 1996 to 2016. Colored groupings illuminate homophilies in Scandinavia (blue), the Arabian Gulf (red), the Balkans (green), Central Asia (orange), and West Africa (purple). Upper Right: The same data, time-averaged by state. Lower: The first principal component, $\mu=\myI+\myE$, of the time-averaged data presented as a map.
\end{figurenotes}
\end{figure}

Given that the time constants are short compared to the centuries-long evolution of the global economy, contemporary observations of the phase, $\phi^i$, must be representative of $\myFP$, conditional upon noise in the form of war, civil war, revolution, natural disaster, or isolation\footnote{Isolation limits the exchange of ideas and goods, both essential components of the global economy.}, for instance. Testing Eq. \ref{eq:Variances}, we estimate $\sigma^2_{\mu,\kappa}$ using HDI as a proxy for $\myE$, and as a proxy for $\myI$ we average five of the World Bank’s six Worldwide Governance Indicators\footnote{We exclude the sixth indicator, Government Effectiveness, because it is materially the same as HDI.} \citep{WWGI}—Control of Corruption, Regulatory Quality, Political Stability and Lack of Violence, Rule of Law, and Voice and Accountability—to form a Worldwide Governance Index, WGI. From annual datasets \citep{UNHDI} and \citep{WWGI},  we compile $3210$ joint observations on $189$ countries between $1996$ and $2016$, shown in the upper left panel of Fig. \ref{fig:RawData}. 
The ratio of variances is $8.2$. Using Eq. \ref{eq:Variances}, we infer $\alpha=0.48$, confirming our conjecture that $\alpha=\myO(0.5)$. The correlation of $\myI$ and $\myE$ has long been taken as causation, but it is in fact a predicted signature of the underlying dynamics.

\begin{figure}
\includegraphics[width=12cm, height=8cm]{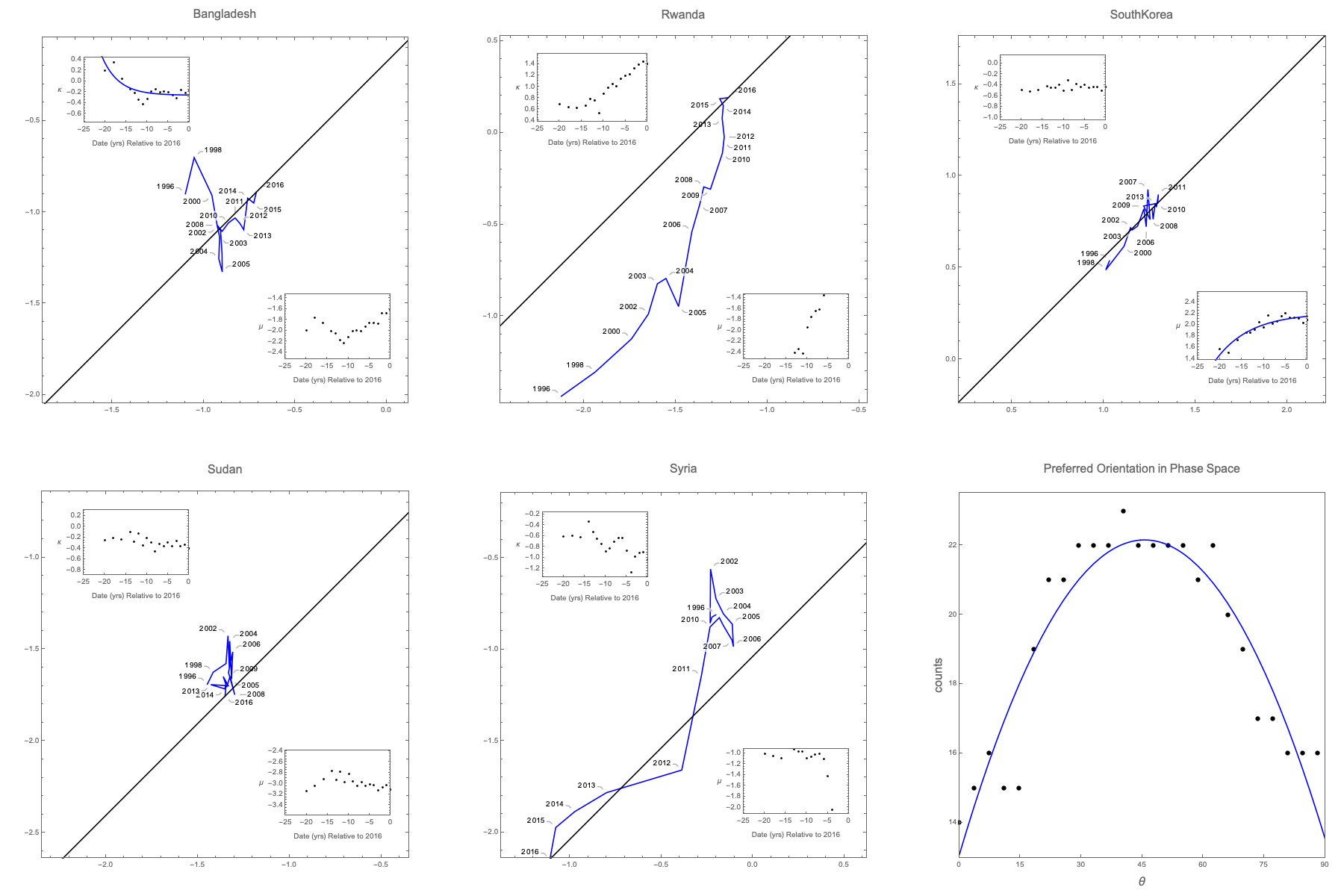}
\caption{Phase space trajectories of individual states.}
\label{fig:Covergence}
\begin{figurenotes}
Bangladesh shows evidence of convergence along $\kappa$ and South Korea along $\mu$. Rwanda and Syria are swept up in long term geopolitical trends, but are not in convergence. Sudan shows evidence of having long since converged to its fixed point. The lower right panel presents the evidence for preferred convergence along the main diagonal. The maximum count at $\theta\!\simeq\! 45^\circ$ validates our equal dissipations assumption  The parabolic fit, $p< 3\!*\!10^{-11}$, illuminates the point.
\end{figurenotes}
\end{figure}

As a final test of this theory, we ask whether it is possible to observe the phase flow to $\myFP$ that is characteristic of the central panel of Fig. 3. We show, in Fig. \ref{fig:Covergence}, the $1996$ to $2016$ phase space trajectories of five states. 
Projecting these on lines of slope $\pm 1$, the $\mu$ and $\kappa$ axes respectively, we extract a time series (shown as insets in the lower right and upper left in each panel) from which we estimate the time constants, $\tau$. 

There is nothing privileged about this period of history. Most states will have long-since gravitated toward their equilibria and their motion in phase space will be stochastic, i.e. Sudan. We expect to see secular motion only in those states caught up in multi-decadal economic or geopolitical trends, for instance Rwanda and Syria. And, in those few that give evidence of relaxation toward an equilibrium, we have no way of knowing whether the inferred equilibrium is ephemeral. 

That said, we fit $\tau$ for each and screen for $p<0.1$, leaving us with $21$ sound estimates of $\tau^i_\mu$ and $\tau^i_\kappa$. Averaging, we get 
\begin{equation} \nonumber
\begin{split}
\bar{\tau}_\mu &=10.5, \sigma=7.5, N=11 \\
\bar{\tau}_\kappa & =\ \ \!6.1, \sigma=4.4, N=10 
\end{split}
\end{equation}
These are consistent with Eq. \ref{eq:Taus} which tells us that $\tau_\mu>\tau_\kappa$. From the ratio of time-constants we infer $\bar{\alpha}=0.26$ and $\bar{\lambda}=0.13$. Given the large variances of the measured $\tau_\mu$ and $\tau_\kappa$, these inferences confirm our \textit{a priori} estimates.

These trajectories also provide us with an independent means to test the validity of our choice of the special case $\gamma=1$. If $\gamma \ne 1$, then the preferred direction of convergence is not along the $\mu$ direction or its conjugate, $\kappa$. Is there a more preferred direction? Rather than project the trajectories along lines of slope $\pm 1$ to extract $\tau_{\mu,\kappa}$, we instead project them along lines of angle $\theta$ and $\theta+90^\circ$, tally the number of sound estimates of $\tau$, and then plot the sum as a function of angle. The maximum of that sum, at or around $\theta=45^\circ$ in the lower right panel of Fig. \ref{fig:Covergence} is consistent with the equal dissipations assumption $\gamma=1$.

\section{Summary}

The large set of humanly devised systems that influence the economy contains elements of two fundamentally different kinds. One, which we call Institutions, $\myI$, varies on a time scale of the economy itself, and the other, which we call Norms, $\myN$, varies much more slowly. Though convention groups these types together\footnote{One notable exception to this convention can be found in the work of Deirdre McCloskey. Equating ethics and norms, she argues that mainly ethics, not mainly law—or what we would call Institutions—is what holds societies together.\citep{Mccloskey2016}}, the dynamical theory embodied in Eq. \ref{eq:Stable2} separates them because $\dot{\myI}\gg\dot{\myN}$. The theory makes several predictions, all of which are borne out. First, the long-term persistence of the distribution of the economic performance of nations follows because the loci of fixed points, $\myFP$, are set by slowly varying exogenous factors and Norms. Second, the distribution of $\myFP$ is expected to have a large principal component along the $\mu=\myI+\myE$ direction. Easily visible in the upper left panel of Fig. \ref{fig:RawData}, the measured value of ratio of variances is $8.2$. Third, regarding the time constants for convergence to $\myFP$, theory predicts $\tau_\mu>\tau_\kappa$ for $0\le\alpha<1$. Subject to the assumptions $\lambda=0.2$ and $\alpha=0.5$, the time constants are $10$ and $3.3$ years, respectively. From the phase space trajectories of $186$ nations over the period $1996$ to $2016$, most of which are noisy, we extract estimates, with large variances, of $\bar{\tau}_\mu=10.5$ and $\bar{\tau}_\kappa=6.1$. Both are comfortably close to the predicted values. Finally, theory says that convergence in phase space proceeds along a preferred direction determined by the ratio of the dissipation constants, $\gamma$. Testing all directions, the lower right panel of Fig.  \ref{fig:Covergence} demonstrates that there is a preferred direction that is consistent with our assumption $\gamma=1$. 

What then does theory tell us about Institutions and Norms? Institutions are the humanly devised systems that promote growth in the economy and vary on the same time scale as economic performance itself. The system equations tell us something else, too. Institutional change keeps pace with the economy. A specific rule, for instance property rights, is not an instance of $\myI$, because a rule does not grow, though a book of rules might qualify. A specific rule, or respect for rules might qualify as a Norm, though whether a specific cultural norm had or has any effect on the fixed points is not something we can say without analysis of materially all possible $\vec{x}$ and $\myN$. Now, consider the infrastructure to support property rights, for instance a network of legislators, courts, and enforcement bureaucracies, each with its buildings and professional staffs, each growing to keep pace with an expanding $\myE$. This infrastructure is $\myI$. Generalizing, $\myI$ is infrastructure of the kind that supports economic growth. And like other forms of infrastructure, it decays to inutility without (expensive) routine maintenance. Norms, $\myN$, are the slowly changing codes of conduct, traditions, convention, and taboos which bear, positively, negatively, or indifferently on $\dot{\myE}$ and $\dot{\myI}$. In the same way that not all infrastructure bears on $\dot{\myE}$, neither do all Norms bear on $\dot{\myE}$ or $\dot{\myI}$. For instance, the rules governing the permissibility of first cousin marriage are Norms, but as to whether those exogamy conventions promote, retard, or remain silent on $\dot{\myE}$ or $\dot{\myI}$, only quantitative analysis will tell.

The system eigenvectors $\mu$ and $\kappa$ define new directions in phase space. Given that $\sigma^2_\mu\gg\sigma^2_\kappa$, most of the system variance is along $\mu$, but what is $\mu$? It is the total economy, the sum of the performance, that is $\myE$, and the infrastructure $\myI$ that supports its growth. In phase space, $\mu$ is the distance measured along the main diagonal whose slope is $+1$. See the upper right panel of Fig. \ref{fig:RawData}. If Norway is an exemplar of social, economic, and political development, or simply order, and Somalia is the opposite, then, $\mu$ measures Social Development or Fortune. Alternatively, $-\mu$ could be said to measure Social Entropy. No analogy is entirely satisfactory.

Names aside, $\mu$ is the proper measure of Development Economics, and as such must be the dependent variable in any modeling or prediction enterprise. The thesis of the New Institutionalism \citep{North1990}, that institutions are the rules of the game and, as such, are independent variables, must be put aside in favor of the first term Eq. \ref{eq:FixedPoints}, which we simplify by exploiting persistence, absorbing constants in $f$ and $g$, and making the assignment $f+g\rightarrow f$, leading to 
\begin{equation} \label{eq:Model}
\mu^i=f(\vec{x}^i,\myN^i)
\end{equation}
In the language of causal inference \citep{Pearl2009}, this equilibrium solution of the dynamical theory is a causal model and as such is the basis for forming interventional and counterfactual queries on data \citep{Pearl2018}. The same may not be said of \textit{ad hoc} statistical models such as Eq. \ref{eq:AJR}.

With the by-state data shown in the upper right panel of Fig. \ref{fig:RawData}, we demonstrate the potential of this $\mu$-theory with a two-factor model, Eq. \ref{eq:TwoFactors}, while simultaneously acknowledging that small models are problematic. Half of the variance of $\sigma^2_\mu$ is explained by $T_{*}$, an extremes-of-temperature variable capturing climate adversity\footnote{$T_{*}$ is country-averaged, mean monthly high temperature averaged over the interval 1900 to 2016. An argument may be made that population-weighting is appropriate, though we do not employ it here.} \citep{CRU325}, and by mean elevation, $h$, capturing geographic adversity.
\begin{equation} \label{eq:TwoFactors}
\mu^i=-0.15\times T^i_{*}-1.5\times h^i +f(\vec{x}^i,\myN^i)
\end{equation}
$N=177$ and the larger of the two $p$-values is $2*10^{-12}$. The coefficient of $T_{*}$ corresponds to a $9.7\%$ loss of gross national income per capita per $^\circ \!C$. The other term corresponds to the same loss per 100 meters increase in mean elevation. Both are probably overestimates owing to omitted variable bias, and in turn, their contribution to the explained variance is probably overstated, too. Coefficients of second order terms, e.g., $T^2_{*}$ which might give evidence of an optimal temperature, are indistinguishable from zero and are not shown. 

Aristotle \citep{AristotlePolitics} wrote of the effects of climate on politics and the economy in the 4th century BCE. In the 14th-century, Ibn Khaldun \citep{IbnKhaldun} wrote more expansively on the subject, and Montesquieu \citep{MontesquieuSpirit} went further yet in the 18th, each describing a mechanism and providing data within the limits of his day. In $2009$, Dell et al. \citep{DJO2009} attributed a $9.5\%$ loss of GDP per capita per $^\circ \!C$ to adaptations not different in spirit from those provided by earlier writers. On the basis of North’s epigram, Robinson et al. \citep{AJR2001} formulate Eq. \ref{eq:AJR} and argue that these observations about climate are spurious misassignments of cause and effect. We do not exaggerate when we say that their work has been influential. However, our analysis demonstrates that the epigram is ill-posed and that Eq. \ref{eq:AJR} is \textit{ad hoc}. It is not causal and it is not a sound basis for causal queries or analysis. Consequently, we reject their argument.

We turn, finally, to $\kappa$, the imbalance between $\myI$ and $\myE$. Though $\kappa$ is both predicted and observed to be small, there is no \textit{a priori} reason to expect it to be zero. We may model $\kappa$ using a corollary to Eq. \ref{eq:Model}, and though the dataset is inherently noisier because $\kappa$ is a difference between two large and independently measured quantities, our bivariate model does attribute $4\%$ of $\sigma^2_\kappa$ to $T_{*}$ ($p<0.01$) and $0\%$ to $h$.

A theory of Development Economics must account for growth and drag in the economy and growth and drag on the sources of growth. We have presented here the simplest of such theories and have shown that it predicts several features of the global economy that were previously unexplained or misconstrued. Being the simplest of such theories, much elaboration and many improvements are possible. Among these, we note: (i) though Eq. \ref{eq:Stable2} applies to one state $i$, we might couple its development with other states $j$ by introducing terms of the form $c_{ij}\myE^j$ and similars for $\myI^j$, their time derivatives, and so forth; (ii) though we have treated $\lambda$ as a constant, we might introduce state-specific drag, $\lambda^i$ , such as would arise from state-specific inefficiencies.

\bibliographystyle{aea}
\bibliography{myrefs}



\end{document}